\documentstyle[prb,aps,epsf]{revtex}

\newcommand{\Ca}{Ca$_{0.83}$CuO$_2$}
\newcommand{\Sr}{Sr$_{0.73}$CuO$_2$}
\newcommand{\Ba}{Ba$_{0.63}$CuO$_2$}

\newcommand{\Li}{Li$_2$CuO$_2$}
\newcommand{\Na}{NaCuO$_2$}
\newcommand{\LaSr}{La$_{1.66}$Sr$_{0.34}$CuO$_4$}

\begin{document}

\draft

\title{Localisation of doped holes in edge-shared CuO$_2$ chain cuprates: consequences for 
dynamic spectral weight transfer }

\author{Z.\ Hu, S.-L.\ Drechsler, J.\ M\'alek\cite{mal}, H.\ Rosner, R.\ Neudert, M.\ Knupfer,
M. S.\ Golden, and J.\ Fink}
\address{Institute for Solid State Research, IFW Dresden, P.O.\ Box
270016, D-01171 Dresden, Germany}

\author{J.\ Karpinski}
\address{Laboratorium f\"ur Festk\"orperphysik, ETH Z\"urich
CH-8057 Z\"urich, Switzerland}

\author{G.\ Kaindl}
\address{Institut f\"ur Experimentalphysik, Freie Universit\"at Berlin
Arnimallee 14, D-14195 Berlin-Dahlem}

\author{C. Hellwig and Ch. Jung}
\address{BESSY GmbH, Albert-Einsteinstr 15, D-12489 Berlin}

\twocolumn[\hsize\textwidth\columnwidth\hsize\csname@twocolumnfalse\endcsname
\maketitle

\begin{abstract}
We present a  joint experimental and theoretical study of the 
electronic structure of the cuprate chain systems 
A$_{1-x}$CuO$_2$ (A=Ca,Sr,Ba), as measured using O-K and Cu-L$_3$  
x-ray absorption spectroscopy.
The doping-dependent behaviour in these systems is radically different to that in 
conventional 2D cuprate networks formed from corner sharing CuO$_4$ plaquettes,
and follows from the strongly suppressed inter-plaquette hybridisation resulting from the 
90$\raisebox{1ex}{\scriptsize o}$ Cu-O-Cu interaction pathway in the chain systems.
Spectroscopically, this results in (a) a classical mixed-valent scenario whereby the 
different final states in the Cu-L$_3$ spectra can be used directly to 'read-off' the Cu
valency and (b) a drastic reduction in the dynamic spectral weight transfer from
the upper Hubbard band to the low energy scale in the O-K spectra.
The final picture emerges of localisation of the doped holes, with the chain then comprised
of a mixture of pure Cu(II)O$_4$ and Cu(III)O$_4$ plaquettes.
\end{abstract}

]

\vspace{1cm}


Low dimensional cuprate systems, such as chains and ladders, gained interest originally 
in the context of their being model systems for the high temperature cuprate superconductors.
It soon became clear that such quasi-1D systems offer a rich correlated electron physics of their
own,\cite{Motoyama,Neudert} with the experimental observation of the separation of the spin and
charge degrees of freedom being one of the highlights so far.\cite{Kim}
\par
A second area to catch the imagination of researchers worldwide has been that of the spin ladders.
Here, superconductivity  has been predicted theoretically for a
doped ladder,\cite{Rice} but has not been observed so far in  pure ladder compounds.

Consequently, the discovery of superconductivity under pressure in
Sr$_{0.4}$Ca$_{13.6}$Cu$_{24}$O$_{41+\delta}$ [\onlinecite{Uehara}] - which has
no CuO$_2$ planes, but rather a mixture of linear CuO$_2$ chains (built up of edge-shared CuO$_4$
plaquettes) and two-leg Cu$_2$O$_3$ ladders - has sparked still further interest in the properties of
the ladders and the chains as individual elements.\cite{Nuecker2000}

In this paper, we present a joint experimental (O-K and Cu-L$_3$ XAS) and theoretical investigation
of the electronic structure of hole-doped edge-sharing CuO$_2$ chains.
The compounds under investigation are {\Ca}, {\Sr} and Ba$_{0.67}$CuO$_2$, all of which
contain CuO$_2$ chains formed of edge-sharing CuO$_4$ plaquettes.
The chains are doped with either 0.33, 0.52 or 0.67 holes per Cu site,
respectively.\cite{Meijer,Shengelaya,Karpinski,Rozhdestvenskaya}
Remarkably, the magnetic susceptibility of {\Ca} and {\Sr} - despite their high hole-doping levels - can be
described within the dimerized alternating spin-1/2 Heisenberg-chain model\cite{Meijer}, with
antiferromagnetic order setting in below 10 K and 12 K for {\Ca} and {\Sr}, respectively.
For the better known CuO$_2$ plane, neither N\'eel nor any other
long-range magnetic order has  been found for doping levels beyond    
0.04 to 0.125 holes per Cu site (i.e.\ beyond the spin-glass and stripe region).
Furthermore, there have been reports of superconductivity with
T$_c=$13 K in the related compound Ba$_2$Cu$_{3-x}$O$_{6-y}$ \cite{Moshkin}, which naturally 
would necessitate the delocalisation of doped holes in  the A$_{1-x}$CuO$_2$ systems.  

\par
The synthesis and the structural analysis of the polycrystalline samples of {\Ca}, {\Sr} and
Ba$_{0.67}$CuO$_2$
have been described previously\cite{Meijer,Karpinski}.
The XAS  measurements were performed at the SX700/II monochromator operated by the
Freie Universit\"at Berlin and the PM5 beamline at the Berliner Elektronenspeicherring f\"ur
Synchrotronstrahlung (BESSY).
In all measurements the non-surface-sensitive fluorescence-yield mode was used.
The energy resolution of the monochromator was set to 280 meV and
600 meV at the O 1$s$ and the Cu 2$p_{3/2}$ thresholds, respectively.
The O 1$s$ data were corrected for the energy-dependent incident flux
and normalized 60 eV above the threshold.
The self-absorption effects were taken into account as described in
Refs.\  \onlinecite{Jaklevic} and \onlinecite{Tröger}. 
Prior to the measurements, the surface of the sintered tablets was scraped
in-situ with a diamond file at a base pressure of 5x10$^{-10}$ mbar.
\par
The simulated Cu-L$_3$ and O-K spectra were calculated with the aid of an extended
five-band $pd$ Hubbard-model as described in Ref. \onlinecite{dagotto94}.
The Cu-L$_3$ and O-K $1s$-XAS spectral density
was calculated by means of the exact diagonalisation of small periodic 
(CuO$_{2}$)$_n$ clusters (with $n$=3,4), following Refs.\ \onlinecite{dagotto94} and \onlinecite{Hybertsen92}.
The calculated data were broadened with Lorentzian functions of width 0.4 eV, in order to 
ease their comparison with experiment.

\par
Fig. 1 illustrates the systematic evolution of the Cu-L$_3$ XAS spectra of {\Ca}, {\Sr} and {\Ba}.
The left panel shows the experimental spectra, whereas the right panel contains the results of
the extended five-band $pd$ Hubbard model referred to above.\cite{Footnote_parameters}
Also shown are data from {\Li} and {\Na}, which serve as formally divalent and trivalent
reference systems for the edge-shared CuO$_2$ chain geometry.
The spectral signatures of {\Li} and {\Na} are simple, single component white-lines at 930.7 and
932.4 eV, respectively.
\par
Fig. 1 shows clearly that the edge-shared chain systems with non-integer formal valence 
exhibit a Cu-L$_3$ XAS spectrum resembling a superposition of scaled versions of the divalent
and trivalent reference systems.
Based upon the excellent agreement observed between experiment and theory in Fig. 1,
we can assign the features as follows. 
For {\Li} the single peak is described by the 2\underline{$p$}3$d$$^{10}$ configuration
(where 2\underline{$p$} denotes the core hole). 
The dominant final state configuration representing the main peak at 932.4 eV in NaCuO$_2$ is described by 
2\underline{p}3d$^{10}$\underline{$L$}. 
Consequently, for {\Ca}, {\Sr}, and {\Ba} the low and higher energy features result from
2\underline{$p$}3d$^{10}$ and 2\underline{$p$}3d$^{10}$\underline{$L$} final states, respectively.
Furthermore, the clear energetic alignment with the divalent and trivalent reference systems
confirms that the features in the Cu-L$_3$ XAS of the intermediate doped chains signal the existence
of Cu(II)O$_4$ and Cu(III)O$_4$ plaquettes.
In line with this picture, the spectral  weight of the higher energy peak increases in line with the formal
valence on going from {\Ca} to {\Sr} and further to {\Ba}.

To put these results in context, we show in Fig. 2 a comparison of the Cu-L$_3$ XAS spectra of two 
formally trivalent systems - {\Na} and LaCuO$_3$ (the latter from Ref. \onlinecite{Mizokawa}) - and two formally
Cu 2.33+ valent systems - {\Ca} and {\LaSr} (the latter from Ref. \onlinecite{CTTJ}).
The enormous differences between the spectra of compounds with the same formal valency is a
signal of the crucial role played by the geometry of the Cu-O network, not only in the Cu 2p core
level photoemission spectra,\cite{VeSa,Boeske1,Boeske2} but also in Cu-L$_3$ XAS \cite{Mizokawa}.
This point is worth further explanation.
\par
In the edge-sharing CuO$_2$ chain systems the Cu-O-Cu interaction
pathway is essentially
90$\raisebox{1ex}{\scriptsize o}$, which strongly suppresses the inter-plaquette hybridisation. 
For the undoped system {\Li}, this has been observed to lead to a behaviour analogous to that
of an isolated CuO$_4$ plaquette.\cite{Uchida,XPS_EELS}.
This extreme localisation means that the spectroscopy of such cuprates is reminiscent of the
situation in some mixed valent 4f compounds in which the intensity ratio of the features at
the (Ln)-L$_3$ thresholds can be used to determine the average valence.\cite{HUMeyer}  
In the case of the CuO$_2$ chain systems, due to the geometrically mediated switching-off of the inter-plaquette hybridisation,
one observes a clear feature at 932.4 in the Cu-L$_3$ spectra whose weight scales with the
formal copper valence.
\par
In contrast, the systems containing strong 180$\raisebox{1ex}{\scriptsize o}$
Cu-O-Cu interaction pathways (LaCuO$_3$ and {\LaSr}), support significant inter-plaquette hopping, thus resulting in an
increase in the effective covalency.
This has two effects.
Firstly, it reduces the contribution of the bare 3$d^n$ configuration in the ground state.\cite{HuCP,HuNi}
Secondly, the inter-plaquette hybridisation offers the system further opportunity to screen the
copper core hole involving the transfer of electron density from ligand levels further away from the
core ionised site.
As a result of this, the Cu-L$_3$ XAS spectrum of trivalent LaCuO$_3$ exhibits a feature at the same
energy as is usually associated with divalent copper, but which actually stems from a trivalent final state
with additional (non-local) screening.\cite{Mizokawa}
The same behaviour is seen for the formally 2.33+ valent cuprates: the edge shared chain exhibits
two clear final state peaks, whereas the interplaquette hopping prevalent in the 2D CuO$_2$-plane
system results in a mere asymmetry at higher energy.\cite{Footnote1}
\par
The spectroscopic consequences of the geometry-induced localisation are not restricted to the Cu-L$_3$
spectra, as is illustrated by the O-K spectra shown in Fig. 3.
The pronounced peak directly above the absorption onset at 530.2 eV in the undoped {\Li} system is related to transitions
into O 2$p$ states hybridized with the Cu$3d$ upper Hubbard band (UHB).\cite{Neudert_Li_XAS}
The pre-edge peak in the O 1$s$ XAS spectrum of the trivalent cuprate {\Na}, which in the context of the HTSC would be
assigned to the doped hole states, is down-shifted by 1.3 eV with respect to the Cu(II) UHB, since with increasing Cu
valence the covalence increases.\cite{Mizokawa,HuCP,HuNi}
\par
In conventional, corner-shared 2D-planar systems which have been hole-doped, two pre-edge peaks are observed
on O-K XAS.
With increasing doping level the UHB quickly loses its intensity, whereas the hole peak gains in intensity and shifts to
lower energy as the doping progresses.\cite{Finkreview}
As can be seen from Fig. 3, the intermediate-doped edge-shared cuprate chains present a different picture.
Firstly, here, as was the case for the Cu-L$_3$ spectra, the two features remain at the same energies as their counterparts in the 
formally Cu(II) and Cu(III) reference systems. 
Secondly, the {\it rate} at which the UHB spectral weight is tranferred to the low energy scale is drastically reduced in these compounds
compared to their corner-sharing 2D cousins.
This is much more than an XAS detail, as this spectral weight transfer is an important characteristic of doped effective Mott-Hubbard 
insulators,\cite{Finkreview} and has been extensively studied in the corner-shared, layered cuprates both experimentally
\cite{CTTJ,CTSE,Romberg} and theoretically \cite{Eskes,Meinders}.
In the corner-shared systems such as La$_{2-x}$Sr$_x$CuO$_4$  [Refs. \onlinecite{CTTJ,Romberg,CTSE}] and the two-leg ladder system 
La$_{1-x}$Sr$_x$CuO$_{2.5}$  [Ref. \onlinecite{TMO}], this effect - expressed in the doping-mediated destruction of the UHB intensity - 
has been found to be supralinear with $x$, and has thus been dubbed dynamical spectral weight transfer or DSWT \cite{Eskes,Meinders}.
For example, in the case of La$_{2-x}$Sr$_x$CuO$_4$ for x=0.15, the UHB is scarcely visible \cite{CTTJ,CTSE} and for the analogous
doping level in the two-leg ladder, the UHB-peak has already lost more than 40 $\%´$ of the spectral weight it had for $x$=0.\cite{TMO}
\par
However, as can be clearly seen in Fig.\ 3 for an even greater doping level of 0.34 in {\Ca} we still find $\sim$  67$\%$ spectral intensity
of the UHB compared with {\Li}, the latter acting as a reference point for the undoped case.
Even {\Ba}, which has a formal doping level of 0.67 holes per Cu, still shows a clear indication of an UHB feature in the O-K XAS.
The suppression of the DSWT is a direct result of the extremely small inter-plaquette hopping along the CuO$_2$ chains, and is 
well reproduced by the extended five-band $pd$ Hubbard model,[\onlinecite{Footnote_parameters}] as can be seen from the good 
agreement between experiment (left panel) and theory (right panel) in Fig. 3.
In addition, this knowledge has played an important role in our understanding of recent polarisation-dependent XAS data from the 
(La,Y.Sr)14' systems\cite{Nuecker2000}, which contain, apart from the 2-leg ladders,
highly doped CuO$_2$ chains.

\par
Fig. 4 summarises the Cu-L$_3$ and O-K core level excitation data of the edge-shared chain systems.
In the upper panel, the mean Cu valencies of Li$_2$CuO$_2$ and the three doped chain compounds under investigation, derived from the 
intensity of the Cu(II) and Cu(III) signals in the Cu-L$_3$ XAS are compared to the formal valence derived from the stoichiometry,
yielding an almost perfect agreement.
This emphasises the unique nature of these cuprate chains, as, to the best of our knowledge, they represent the first examples
of a mixed valent cuprate family, in which core level spectroscopy allows one to 'read off' the mean Cu valence merely by comparing
the intensity ratio of the double-peaked Cu-L$_3$ XAS feature.
The lower panel of Fig. 4 shows an analysis of the doping dependent evolution of the spectral weight of the hole peak and the UHB in the O-K XAS
spectra,
whereby the doping dependence of the ZRS and UHB features in {LS} (from Ref. [\onlinecite{CTSE}]) has been included for comparison.
The strong suppression of the DSWT in the edge-shared chain systems is very clear.
clearly indicating the supression of the dynamic spectral weight transfer in these systems.

\par
To summarize, we have presented a joint experimental and theoretical study of the effects of hole doping on the electronic structure of
CuO$_2$ chains built up of edge-shared CuO$_4$ plaquettes.
Analysis of the data, both at the qualitative and quantitative level, illustrates the remarkable consequences of the
edge-sharing geometry of these systems, which results in a 90$\raisebox{1ex}{\scriptsize o}$ 
Cu-O-Cu interaction pathway, thereby essentially switching-off the inter-plaquette hopping.
\par
(1) The system is robbed of the possibility of using non-local processes to screen the core hole in Cu-L$_3$ XAS, thus resulting a
in a classical mixed valence behaviour in which the average Cu valence can be simply extracted from the relative intensity of the 
2\underline{$p$}3d$^{10}$ and 2\underline{$p$}3d$^{10}$\underline{$L$} final state features. 
This is an unprecendently clear example of the impact of non-local effects in XAS.
\par
(2) The dynamic transfer of spectral weight from the upper Hubbard band to the low energy scale is strongly suppressed, resulting in the 
observation of the UHB feature in the O-K spectra even up to for a doping level of 0.67 extra holes per Cu.
\par
(3) Together with the observation of Neel order in the {\Sr} and {\Ca} systems,\cite{Meijer} the spectroscopic data presented here provide
clear evidence for the existence of individually identifiable Cu(II)O$_4$ and Cu(III)O$_4$ plaquettes in these materials.

 \vspace {0.3cm}

This work was funded by the Deutsche Forschungsgemeinschaft (Fi439/7-1 and the Graduate College 'Structure and correlation effects
in solids' of the TU Dresden). H.R.\ is grateful to the DAAD for financial support.

\begin{figure}
\caption{
Left panel: Cu-L$_3$ XAS spectra of edge-shared CuO$_2$ chain systems as a function of hole
doping. From the bottom:
{\Li}, {\Ca}, {\Sr}, {\Ba} and {\Na}.
The right panel shows the same spectra simulated using an extended five band $pd$ Hubbard model.
The hole doping level appropriate for each system is indicated next to each theoretical curve.
For details see text.
}
\end{figure}

\begin{figure}
\caption{
Cu-L$_3$ XAS spectra of (from the top) the formally trivalent systems {\Na} and LaCuO$_3$
(from Ref. [\protect\onlinecite{Mizokawa}]),
together with the formally Cu 2.33+ valent systems {\Ca} and {\LaSr} (the latter from Ref. [\protect\onlinecite{CTTJ}]).
}
\end{figure}

\begin{figure}
\caption{
Left panel: O-K XAS spectra of edge-shared CuO$_2$ chain systems as a function of hole
doping. From the bottom: {\Li}, {\Ca}, {\Sr}, {\Ba} and {\Na}.
The right panel shows the same spectra simulated using an extended five band $pd$ Hubbard model.
The hole doping level appropriate for each system is indicated next to each theoretical curve.
For details see text.
}
\end{figure}

\begin{figure}
\caption{
(a) The mean Cu valence (filled triangles) obtained from the Cu-L$_3$ XAS spectra of 
 {\Li}, {\Ca}, {\Sr}, {\Ba} together with the formal valence (filled circles) from the stoichiometry plotted versus the 
hole doping level, $x$.
(b) the relative spectral intensity of the doping-induced hole peak (ZRS, open squares) and the UHB
(filled squares) in the O-K XAS spectra of the same systems, normalized to the intensity of the UHB for undoped
{\Li}.
The decay of the UHB intensity and the growth of the ZRS feature for the system La$_{1-x}$Sr$_{x}$CuO$_4$
(from Ref. [\protect\onlinecite{CTSE}])
is shown as filled and open grey circles for comparison.
}
\end{figure}


\begin{thebibliography}{26}
\bibitem{Footnote1} The vast majority of doped CuO$_2$-plane systems studied with XAS exhibit similar Cu-L$_3$
spectra to that of the La-214 system \protect[\onlinecite{Finkreview}], with the exception of the Hg-cuprates, in which the high energy
feature is visible as a clear shoulder or sub-peak on the main white line. However, as the formal doping level in these
systems is difficult to define, we use the La-214 system as a more clear-cut reference.
\end{thebibliography}
\end{document}